\pdfoutput=1
\documentclass[review]{elsarticle}

\usepackage{lineno,hyperref}
\usepackage{siunitx}
\usepackage{placeins}
\usepackage{booktabs}
\modulolinenumbers[5]

\journal{Advances in Space Research}

\usepackage{natbib}
\setcitestyle{authoryear,open={(},close={)}}

\begin{document}

\begin{frontmatter}

\title{The integration and testing of the Mini-EUSO multi-level trigger system}

\cortext[mycorrespondingauthor]{Corresponding authors}

\author[russia1,russia2]{Alexander Belov\corref{mycorrespondingauthor}}
\ead{aabcad@gmail.com}

\author[italy1,italy2]{Mario Bertaina}
\author[sweden1,sweden2]{Francesca Capel\corref{mycorrespondingauthor}}
\ead{capel@kth.se}

\author[italy1,italy2]{Federico Fausti\corref{mycorrespondingauthor}}
\ead{fausti@to.infn.it}
 
\author[italy1,italy2]{Francesco Fenu}
\author[russia1]{Pavel~Klimov}
\author[italy1,italy2]{Marco Mignone}
\author[italy1,italy2]{Hiroko Miyamoto}
\author{for the JEM-EUSO Collaboration}

\address[russia1]{D.V. Skobeltsyn Institute of Nuclear Physics, M.V. Lomonosov Moscow State University, 1(2), Leninskie Gory, Moscow, 119991, Russia}
\address[russia2]{Faculty of Physics, M.V. Lomonosov Moscow State University, 1(2), Leninskie Gory, Moscow, 119991, Russia}
\address[italy1]{Dipartimento di Elettronica e Telecomunicazioni, Politecnico di Torino,\\ Corso Duca degli Abruzzi, 24, 10129 Torino, Italy}
\address[italy2]{Istituto Nazionale di Fisica Nucleare, sez. di Torino, via P. Giuria,1, 10125 Torino, Italy}
\address[sweden1]{Department of Physics, KTH Royal Institute of Technology, SE-106 91 Stockholm, Sweden}
\address[sweden2]{The Oskar Klein Centre for Cosmoparticle Physics, SE-106 91 Stockholm, Sweden}

\begin{abstract}
The Mini-EUSO telescope is designed by the JEM-EUSO Collaboration to observe the UV emission of the Earth from the vantage point of the International Space Station (ISS) in low Earth orbit. The main goal of the mission is to map the Earth in the UV, thus increasing the technological readiness level of future EUSO experiments and to lay the groundwork for the detection of Extreme Energy Cosmic Rays (EECRs) from space~\citep{Ebisuzaki:2014wka}. Due to its high time resolution of \SI{2.5}{\micro\second}, Mini-EUSO is capable of detecting a wide range of UV phenomena in the Earth's atmosphere. In order to maximise the scientific return of the mission, it is necessary to implement a multi-level trigger logic for data selection over different timescales. This logic is key to the success of the mission and thus must be thoroughly tested and carefully integrated into the data processing system prior to the launch. This article introduces the motivation behind the trigger design and details the integration and testing of the logic. 

\end{abstract}


\end{frontmatter}


\section{Introduction}
\label{sec:INTRO}

The Mini-EUSO instrument is designed for the measurement and mapping of the UV night-time emissions from the Earth and is being developed by the JEM-EUSO Collaboration as a pathfinder for the detection of EECRs from space. The cosmic ray energy spectrum extends over many orders of magnitude in both flux and energy. Notably, at the high energies of EECRs, defined as cosmic rays with energy E~$>$~\SI{50}{\exa\electronvolt}, the flux drops to less than 1~particle/\si{\kilo\metre\squared}/century and the exposure of a proposed detector becomes a critical consideration. Existing ground-based experiments make use of large detector arrays and fluorescence-detecting telescopes in order to capture the particle showers and UV light emitted by the interaction of EECRs with the Earth's atmosphere (for example the Telescope Array Project \citep{Kawai:2008gz} and the Pierre Auger Observatory \citep{Abraham:2004dt}). By taking the fluorescence detection concept into Earth orbit, the exposure of such an experiment to EECRs could be dramatically increased, whilst also allowing for coverage of both hemispheres. Increasing the statistics of these measurements is key to improving our understanding of the astrophysical sources and mechanisms which are capable of producing such high energy particles.

The JEM-EUSO collaboration proposes such space-based experiments including the KLYPVE/K-EUSO~\citep{Panasyuk15} on the Russian Segment of the International Space Station (ISS) and JEM-EUSO (Extreme Universe Space Observatory on board the Japanese Experiment Module)~\citep{Collaboration15, Olinto15}. Mini-EUSO is currently approved by both the Russian (Roscosmos) and Italian (ASI) space agencies and is set to be launched to the Zvezda module of the ISS in early 2018, where it will look down on the Earth from a nadir-facing, UV-transparent window, in order to verify this detection concept. 

\begin{figure}[h]
\centering%
{\includegraphics [width=0.9\textwidth]{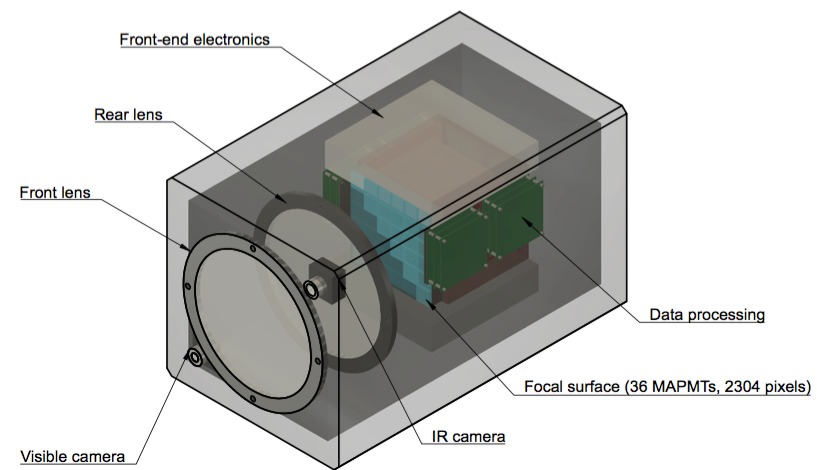}}
\caption{The design of the Mini-EUSO instrument. The main sub-systems are shown: the two double-sided Fresnel lenses, the PDM and the readout electronics. The near infra-red and visible cameras are mounted at the level of the first lens, outside of the optical system. The dimensions of the instrument are 37 $\times$ 37 $\times$ 62 \si{\cubic\centi\metre}.}
\label{fig:minieuso_highlevel}
\end{figure}

Mini-EUSO is made up of three main sub-systems: the Fresnel-based optical system, the Photo-Detector Module (PDM) and the readout electronics. There are also two ancillary cameras installed at the level of the front lens, to provide complementary information in the visible and near infra-red range. The design of Mini-EUSO is shown in Figure \ref{fig:minieuso_highlevel}. The optical system consists of 2 double-sided Fresnel lenses with a diameter of \SI{25}{\centi\metre} allowing for a compact system with a large aperture, ideal for space application \citep{Hachisu:2011vx}. The lenses focus the light onto the focal surface, where it is detected by the PDM. 36 multi-anode photomultiplier tubes (MAPMTs) with UV filters make up the PDM, each with 64 pixels, resulting in a readout of 2304 pixels. Signals are pre-amplified and converted to digital by the SPACIROC3 ASIC \citep{BlinBondil:2014ve}, before being passed to the data processing unit (PDM-DP) for data handling and storage.

The main scientific goal of Mini-EUSO is to produce a high-resolution map of the Earth in the UV range (300 - 400 nm). With a spatial resolution of $\sim$~\SI{5}{\kilo\metre} and a temporal resolution of \SI{2.5}{\micro\second}, Mini-EUSO will present results of unprecedented detail in this range. Such observations are key to the understanding of the detection threshold of EECRs from space, in addition to estimating the duty cycle of future experiments. Although the energy threshold of Mini-EUSO is likely too high to detect EECRs, the instrument's capability to detect such events will be tested by triggering on EECR-like laser tracks produced by ground-based laser systems. Due to its high resolution, Mini-EUSO is also capable of capturing a variety of both atmospheric and terrestrial phenomena, such as transient luminous events (TLEs), meteors, space debris, bioluminescence and anthropogenic lights. The duration of such events varies by 6 orders of magnitude, motivating a multi-level trigger system to maximise the scientific return, given constraints on the duty cycle and data storage. For more details on the Mini-EUSO instrument, see \citet{Mini-EUSO} (in submission).

TLEs are upper-atmospheric events which occur above areas of thunderstorm activity \citep{Pasko:2011ev}. Imaged for the first time in 1989, as described in \citet{Franz:1990cu}, the term TLE is used to describe a wide range of phenomena which are broadly classified into 3 main groups: blue jets, sprites and elves. Blue jets and sprites are more localised, column-like events which can occur in groups and have a spatial extent of a few \si{\kilo\metre} when viewed from above. Elves are very fast flashes which form large halos, extending over \SI{300}{\kilo\metre}. Generally speaking, the timescales associated with TLE events are between $\sim$~\SI{100}{\micro\second} and $\sim$~\SI{1}{\milli\second}, with rise times on the scale of \SI{10}{\micro\second}. TLEs are luminous in the UV and occur with a high frequency, making it important to study such events due to their contribution to the EECR-signal background. In addition, Mini-EUSO will be able to provide high-time-resolution imaging of these events which could contribute to a better understanding of their structure and development in relation to the associated thunderstorms. The Atmosphere-Space Interactions Monitor (ASIM, \citet{Neubert:2009ih}) is designed for the study of atmospheric phenomena, including TLEs, and will operate on-board the ISS during the mission of Mini-EUSO. A specific trigger level for the study of TLEs has been implemented in Mini-EUSO to capture these events and will allow for the joint study of TLEs by both instruments. 

\section{The trigger algorithm}

\subsection{The digitised data path}
\label{sec:signal path}  

\begin{figure}[h]
\centering%
{\includegraphics [width=\textwidth]{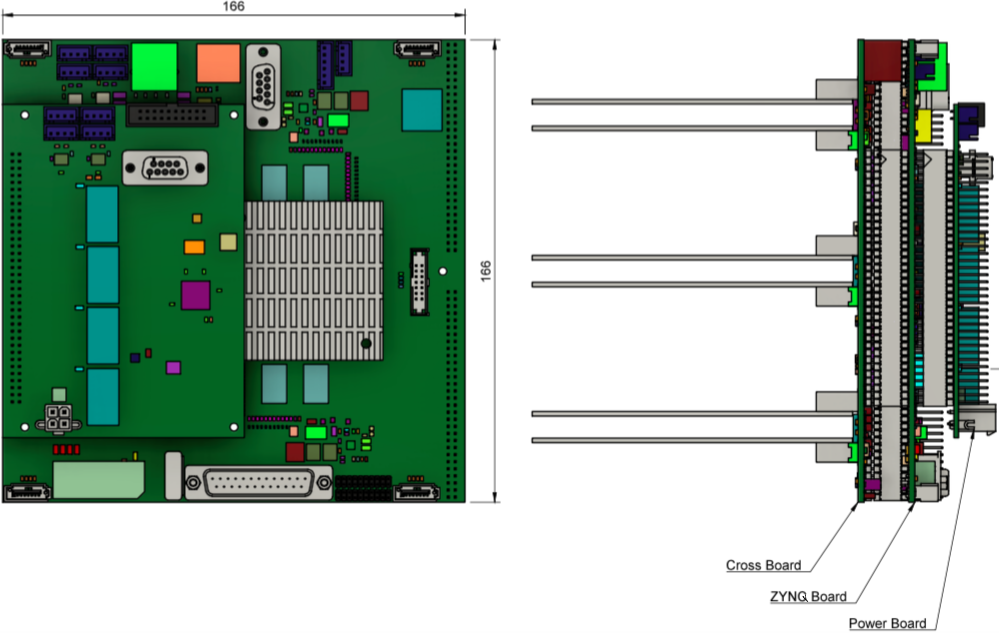}}
\caption{The PDM-DP is shown with dimensions in mm. The 3 separate boards can be seen with the mechanical support for the SPACIROC3 ASICs on their left. }
\label{fig:pdm-dp}
\end{figure}

Photons focused onto the Mini-EUSO focal surface are detected by the 36 MAPMTs, each with 64 pixels (Hamamatsu R11265-M64). Each PMT is read out by a 64 channel SPACIROC3 ASIC (AMS 0.35 $ \mu$m SiGe) operating in single photon counting mode. This data is digitised for each acquisition window of \SI{2.5}{\micro\second} which is referred to hereafter as a gate time unit (GTU), for a data sampling rate of \SI{400}{\kilo\hertz}. The output of the SPACIROC3 ASIC is then passed to the PDM data processing, or PDM-DP. The PDM-DP consists of 3 boards, the cross board, the Zynq board and the power board, as shown in Figure \ref{fig:pdm-dp}. 

\begin{figure}[h]
\centering
{\includegraphics [width=\textwidth]{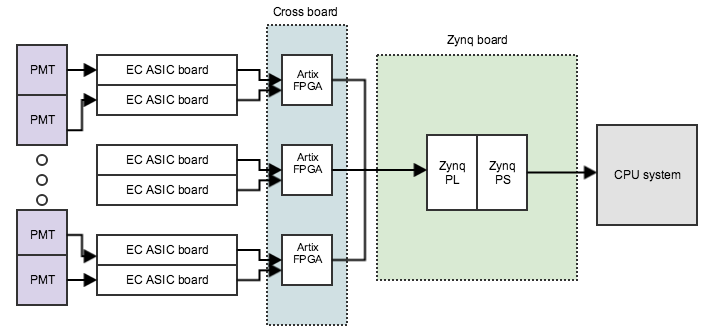}}
\caption{A schematic representation showing the digitised data path.}
\label{fig:dataflow}
\end{figure}

The cross board contains 3 synchronised Xilinx Artix7 FPGAs which perform data gathering from the ASICs, pixel mapping and data multiplexing. Data is output from the cross board in a 48 $\times$ 48 pixel format transferred at \SI{200}{\mega\hertz}, using a \SI{100}{\mega\hertz} double data rate. The Zynq board interfaces to the cross board and contains a Zynq XC7Z030 system of programmable logic (PL) Xilinx Kintex7 FPGA, with an embedded dual core ARM9 CPU processing system (PS). The Zynq board does the majority of the data handling including data buffering, configuration of the SPACIROC3 ASICs, triggering, synchronisation, and interfacing with the separate CPU system for data storage. In addition to these tasks, the high-voltage control to the PMTs is also taken care of by the Zynq board. The power board provides the necessary voltages to the system. Figure \ref{fig:dataflow} summarises the digitised data path.

\subsection{The multi-level trigger}
\label{sec:Trigger}
The Mini-EUSO trigger logic is implemented in VHDL inside the PL of the Zynq Board and consists of two levels, level 1 (L1) and level 2 (L2), that work with different time resolutions. Each level is dedicated to a specific category of events that will be observed by Mini-EUSO. The motivation behind the trigger algorithm is to capture different events of interest on short timescales, but also to provide continuous imaging on slower timescales as Mini-EUSO orbits around the Earth. In order to achieve this efficiently, 3 different types of data are stored, each with different time resolution.

\begin{figure}[h]
\centering%
{\includegraphics [width=0.9\textwidth]{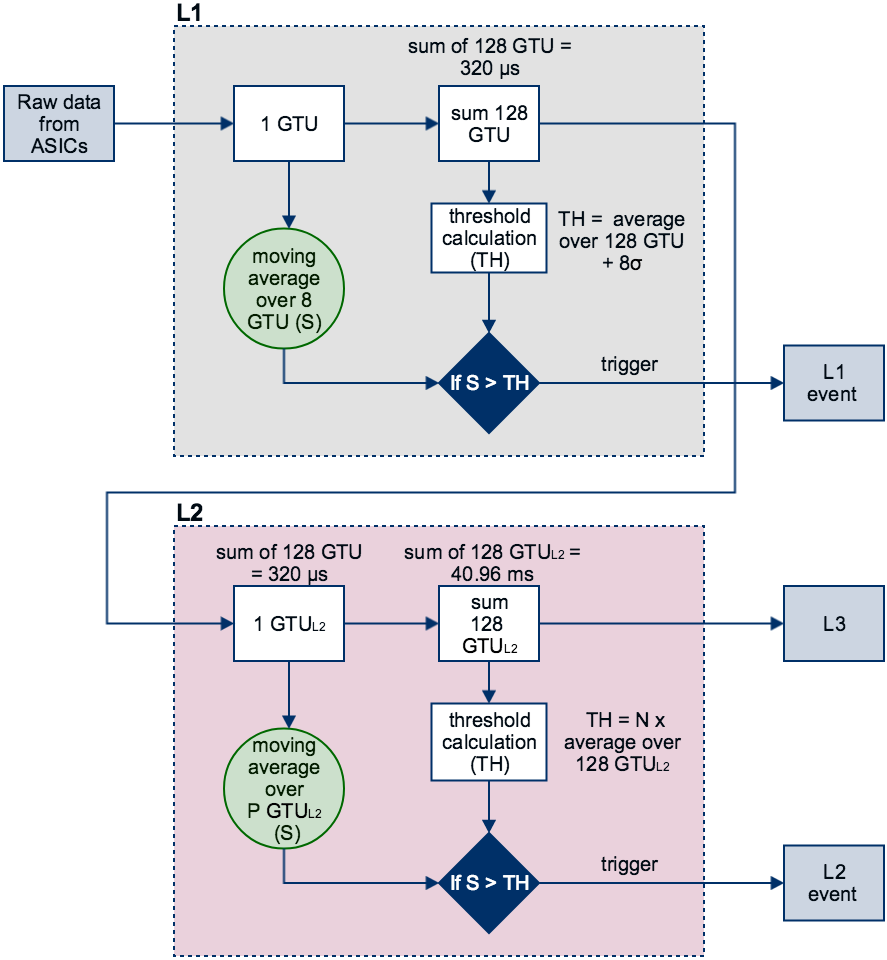}}
\caption{A block diagram summarising the trigger logic. Top: L1, bottom: L2. The trigger outputs 3 separate types of data with time resolutions of \SI{2.5}{\micro\second}, \SI{320}{\micro\second} and \SI{40.96}{\milli\second}. }
\label{fig:L1andL2}
\end{figure}

The L1 trigger gives data with a time resolution of \SI{2.5}{\micro\second} and looks for signal excess on a timescale of \SI{20}{\micro\second}, as this corresponds to the timescale of EECR-like events. Each pixel is considered as independent, motivated by the fact that its field of view at ground is $\sim$~\SI{5}{\kilo\metre}, so light takes at least $\sim$~\SI{20}{\micro\second} to cross one pixel. Pixel signal is integrated over 8 consecutive GTUs and compared with the background level, determined by averaging over 128 GTU, to look for an excess. If the signal is $8\sigma$ above background, the event is triggered, the whole focal surface is read out and a packet of 128 GTU is stored, centred on the trigger. In addition to this, the data integrated over 128 GTU (\SI{320}{\micro\second}) in order to set the background level is then passed to the L2 trigger. 

The L2 trigger receives the integration of 128 GTU (=1 L2 GTU, or GTU$_{L2}$) as input from the L1. It operates with a similar logic, but with a time resolution of \SI{320}{\micro\second}, well-suited to capturing fast atmospheric events, such as TLEs and lightning, which have timescales from $\sim$~\SI{100}{\micro\second} up to $\sim$~\si{\milli\second}. Background is set by integrating 128 GTU$_{L2}$, which is also stored as the level 3 (L3) data, or 1 GTU$_{L3}$. An L2 trigger occurs when the signal in P GTU$_{L2}$ is greater than N times the background level, and the event is stored. These key parameters can be altered in-flight to optimise the trigger performance. Following offline optimisation of the trigger using simulations (as described in Section \ref{l2_test_esaf}), the default values are N~=~4 and P~=~8. The background is determined as a sum of the pixel counts over 128 GTUs, which is then rescaled on P GTUs as shown in Figure \ref{fig:L1andL2}.

After the accumulation of 128 GTU$_{L3}$, or every \SI{5.24}{\second}, all stored events from L1, L2 and L3 data are transferred to the CPU for formatting and storage on the disk. If no L1 or L2 events are triggered, instead the last 128 GTU or 128 GTU$_{L2}$ present on the overwritten buffer are read out. In this way, a continuous and controlled readout is achieved with a resolution of \SI{40.96}{\milli\second} whilst also capturing interesting events at faster timescales. This \SI{40.96}{\milli\second} ``movie'' will be used to search for meteors, space debris and strange quark matter using offline trigger algorithms, as well as for the mapping of the Earth in UV. The L1 and L2 trigger algorithm is summarised in Figure \ref{fig:L1andL2}. 

L1 and L2 thresholds are set to trigger, on average, at a rate lower than 1 event per \SI{5.24}{\second}. Assuming that 3 bytes/pixel are recorded, the presented trigger algorithm gives a data readout of 507 kB/s. Assuming an optimistic duty cycle of 50\%, this results in a data storage requirement of 660 GB/month. Assuming some ancillary data from the camera and housekeeping systems, it is still reasonable to estimate a maximum data output of 1 TB/month. 

The L1 trigger defined for Mini-EUSO is based on the same concept of observing a signal excess relative to the UV background level as that of the trigger for future large-scale EUSO instruments, such as JEM-EUSO. However, the main difference in the implementation is that 1 single pixel of Mini-EUSO has a field of view (FoV) slightly larger than the FoV of a whole MAPMT in JEM-EUSO. Therefore, in Mini-EUSO, each pixel is considered independent and the signal is integrated over 8 GTUs, which is the minimum time that the signal will stay within 1 pixel. In contrast, the JEM-EUSO trigger system will utilise a 3~$\times$~3 pixel array to search for signal excess on timescales of 5 GTU, optimised for the observation of typical EECR signal development by a larger scale instrument. Notably, Mini-EUSO is the first implementation of a specific trigger level dedicated to ``slower'' events (i.e.~$>$~\si{\micro\second} scale) in a pathfinder mission for JEM-EUSO. The second level trigger of the future JEM-EUSO system will be dedicated to further filtering in the EECR signal time domain.

\section{Verification of the trigger algorithm}
\label{sec:testalg}

Prior to the implementation of the trigger algorithm in hardware, the logic has been tested extensively using both simulated data and data taken at the TurLab facility. 

\subsection{L1 trigger tests at TurLab}
\begin{figure}[ht]
\centering
\includegraphics[width=0.7\textwidth]{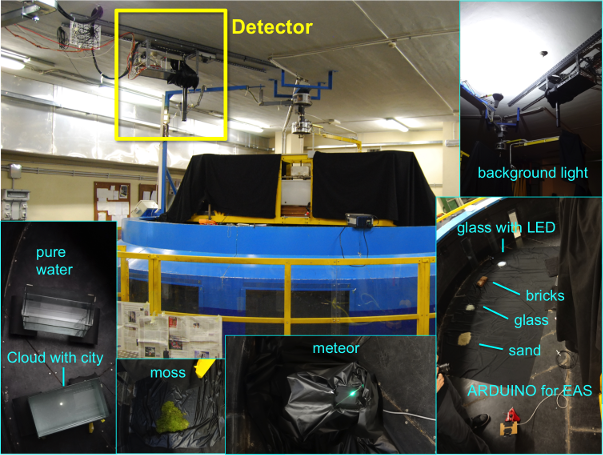}
\caption{The TurLab rotating tank. The black tube on the ceiling shows the
collimator of the experimental setup used to mimic the Mini-EUSO telescope.
Light sources and materials used to mimic other UV sources are also shown.}
\label{fig:vasca}
\end{figure}

The EUSO@TurLab project is an ongoing activity aimed to reproduce atmospheric and luminous phenomena that the JEM-EUSO and EUSO style telescopes will observe from Earth orbit \citep{Bertaina:2015hha}. TurLab is a laboratory equipped with a \SI{5}{\metre} diameter rotating tank  and located \SI{15}{\metre} below ground level. Therefore, without artificial illumination, the room is darker than the night sky by several orders of magnitude. The EUSO@TurLab project makes use of the TurLab rotating tank with a series of different light configurations to reproduce the UV emission of the Earth. The Mini-EUSO detector is represented by one elementary cell (EC) unit of 4 MAPMTs and the necessary readout electronics. The detector is suspended from the ceiling and looks down on the rotating tank to mimic the observation from orbit (see Figure~\ref{fig:vasca}).

The capability of controlling the tank rotation speed (\SI{3}{\second} - 20 minutes per turn) allows for the reproduction of events of different duration and spatial extent, as seen from ISS, with the same configuration. 

Vital to the testing of the trigger algorithm in this setup is the choice and variety of light sources. There are two types of light source:
1) direct light emitting sources;
2) materials reflecting ambient light.
A range of different light sources are used, with the intent of
reproducing different kinds of phenomena:
a) LEDs inside tubes of different dimensions, in order to reproduce extended
intense light directly pointing towards the MAPMT, thus representing urban areas;
b) an oscilloscope generating Lissajous curves for events such as meteors;
c) LEDs driven by a pulse generator for fast luminous events such as lightning;
d) LEDs or optical fibers driven by Arduino for light pulses with \si{\micro\second} duration.
A more detailed discussion of the setup is reported in~\citet{Bertaina:2015hha}.

For these measurements in EUSO@TurLab, the apparatus consists of one fully-equipped EC unit, similar to those used in Mini-EUSO, with a 1 inch focusing lens (\SI{50}{\centi\metre} focal length) placed directly in front of the MAPMTs. A test board is used to retrieve data from the EC ASICs and a LabView program is used for data acquisition. The main differences between the TurLab setup and Mini-EUSO are that data is acquired in packets of 100 GTU instead of the nominal 128 GTU, and the system has a $\sim$~\SI{50}{\milli\second} delay between two consecutive acquisitions of 100 GTU. This condition slows down the measurements and introduces artificial discontinuities in the recorded light between two acquisitions. 200 simulated data packets were added between two experimental packets in order to smooth out such discontinuities. In this way, it was possible to extend the 8.2$\times$10$^5$ GTUs collected in around \SI{7}{\minute} of rotation, to a total number of 1.6$\times$10$^8$ GTUs used to test the L1 trigger offline.

\begin{figure}
\centering
\makebox[\textwidth][c]{\includegraphics[width=1.5\textwidth]{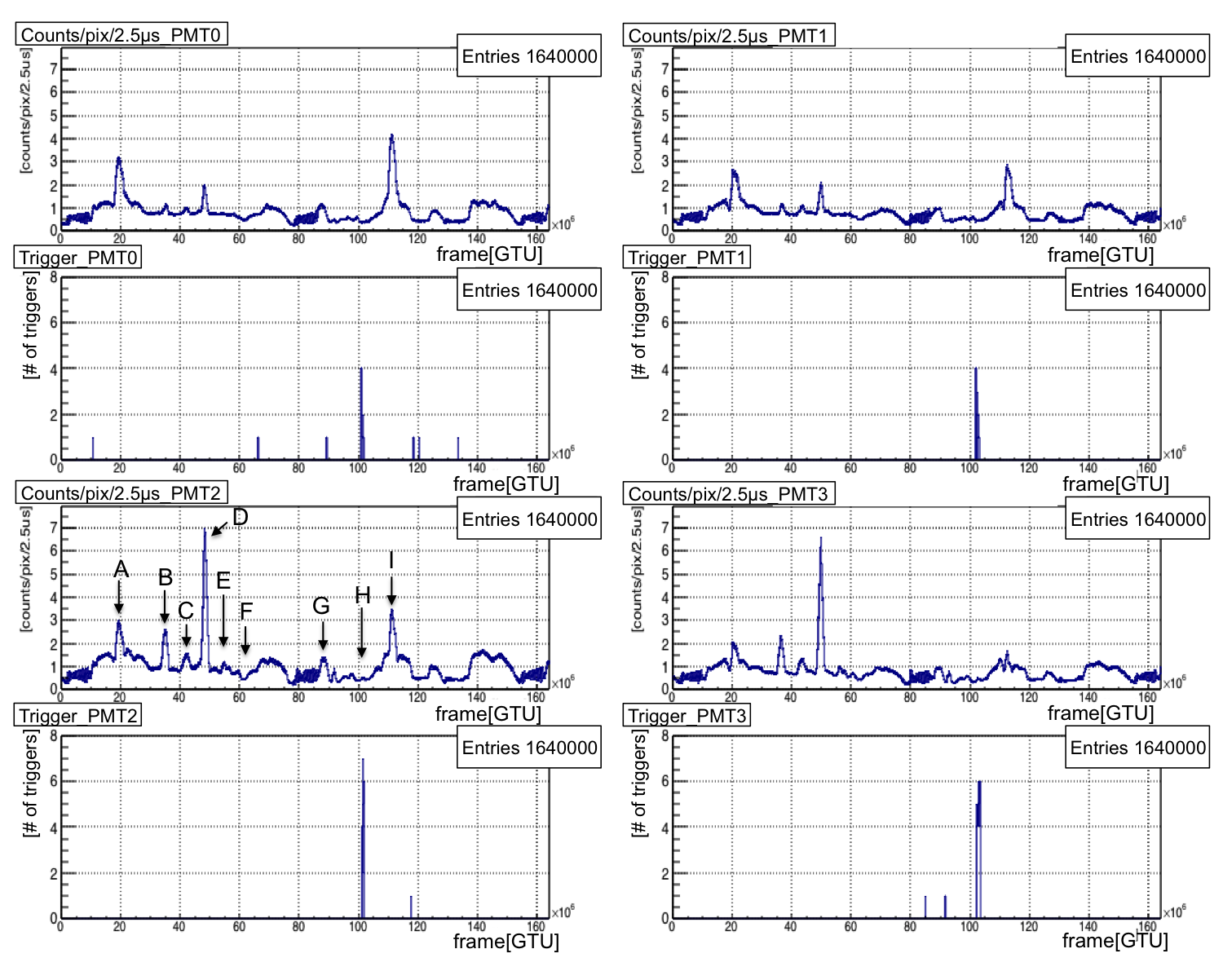}}
\caption{The figure is divided into 4 different blocks. In each block the top plot shows the average number of counts per pixel, normalised at PMT and packet level, as a function of time for one PMT. The bottom plot indicates the time of L1 trigger activation, with the y-axis showing the number of triggers. See the text for further details.}
\label{fig:L1trigger-TurLab}
\end{figure}

Figure~\ref{fig:L1trigger-TurLab} shows an example of the performance of the trigger logic 
described in Section \ref{sec:Trigger} for one EC unit. 
The figure is divided into 4 different blocks. In each block
the top plot shows the average number of counts per pixel, normalised at the PMT and packet level,
as a function of time for one PMT, while the bottom plot indicates the time when the L1 trigger was activated. The different letters (from A to I) in the plot of PMT 2 indicate different types
of light surface or reflective source present in the tank, which are responsible
for a different signal seen by the PMTs. The same pattern is apparent
in all 4 PMTs, but with different intensities and slightly shifted in time 
due to the movement of the tank and the size of the light source. 

Pictures of these sources are displayed in Figure~\ref{fig:vasca}. A represents clouds; B and D represent the response to ground glass in which D looks brighter because it is glass illuminated by an LED; C, E and F are the reflections from sand, brick and moss, respectively; G
is due to meteor-like signals; H is due to an Arduino-emulated cosmic ray and I to the reflection
of clear water. The Arduino event looks quite dim compared to other signals
because the track is limited to a few pixels, therefore, it is almost
overwhelmed by the total number of counts in the PMT.

Despite the presence of several light sources of different
intensity, duration and extension, most of the triggers occur in coincidence with the Arduino EECR-like signal transit in the field of view of the telescope for all four PMTs. 
The rate of spurious triggers re-scaled to one full PDM is $\sim$~\SI{0.2}{\hertz}, which is compatible with the acquisition logic. The rate of the Arduino events was not controlled as the main purpose of these initial tests was to verify the ability of the trigger logic to avoid light signals which are not EECR-like, and this was indeed achieved. Regarding the ability of the trigger to capture EECR-like events, the trigger efficiency has been tested both via ESAF simulations and through hardware tests as detailed in Sections \ref{l1_test_esaf} and \ref{l1_test_hw}.

This data from TurLab was also used for online testing of the VHDL code implemented in FPGA, and the same results were obtained. These results demonstrate that the L1 trigger is sensitive to the presence of EECR-like light signals. 

\subsection{L1 trigger tests with ESAF}
\label{l1_test_esaf}
The main objectives of the TurLab tests were the verification of the capability of the L1 trigger logic and the optimisation of the trigger thresholds with variations of light intensity. This is important in order to keep the rate of false triggers at an acceptable level. The logic demonstrated the capability of recognising and triggering on EECR-like signals. Events of longer duration such as meteors, city lights, clouds, etc. do not generate triggers, as required.

In order to evaluate the trigger performance for EECR observation, simulations using the ESAF code were 
performed. The EUSO Simulation and Analysis Framework (ESAF) \citep{Berat10} is currently used as the simulation and analysis software for the JEM-EUSO and its pathfinder missions. ESAF performs the simulation of the shower development, photon production and transport in the atmosphere, and detector simulations for optics
and electronics. Furthermore, algorithms and tools for the reconstruction of the shower properties are included in the ESAF package ~\citep{Bertaina:2014fk}. Recently,  the Mini-EUSO mission configuration was implemented in ESAF, including the L1 trigger logic, in order to assess its performance.

\begin{figure}[h]
  \centering    
      \makebox[\textwidth][c]{
  	\includegraphics[width=0.8\textwidth]{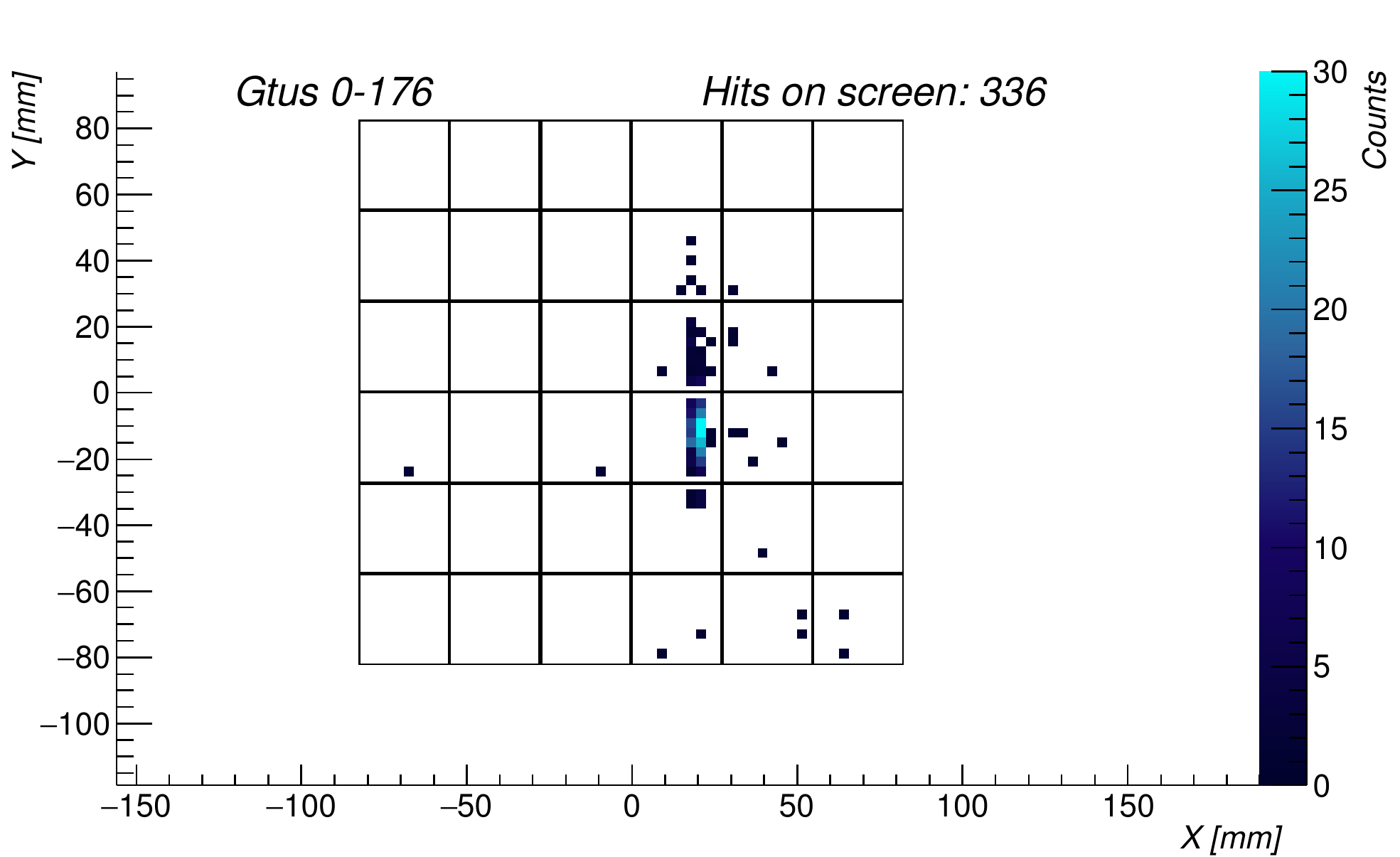} 
    	\includegraphics[width=0.8\textwidth]{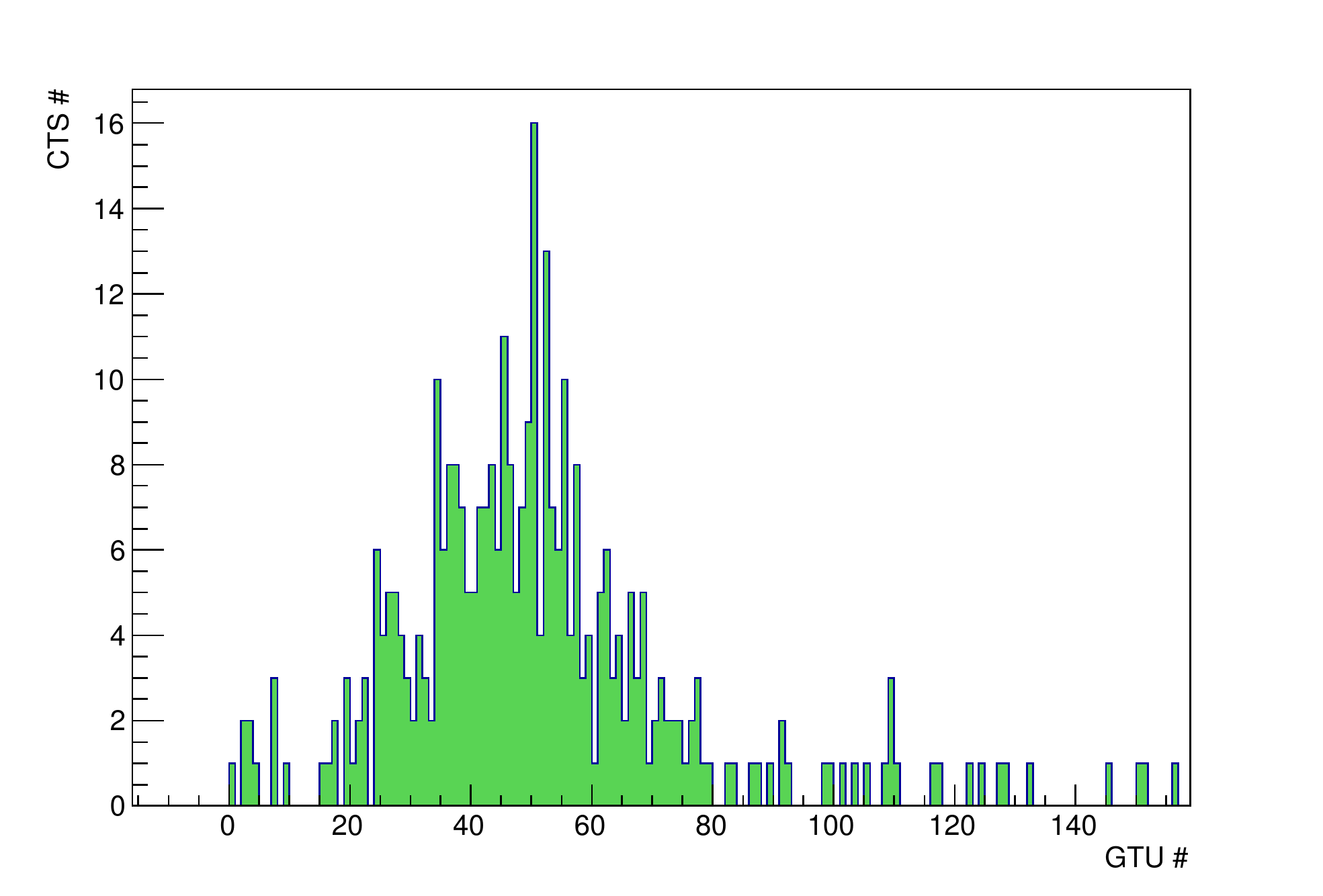}}
  \caption{Left: Photon counts observed in the Mini-EUSO focal surface for a simulation of a $E$ = \SI{1e21}{\electronvolt} event with an inclination of $80^{\circ}$ to the nadir (background is not included in the simulation). Such an event would be on the threshold of detection for Mini-EUSO. Right: Light curve for the same event. The x-axis shows time in units of GTU (1 GTU = \SI{2.5}{\micro\second}).}
  \label{fig:eecr}
 \end{figure}

\begin{figure}[h]
  \centering
  \includegraphics[width=\textwidth]{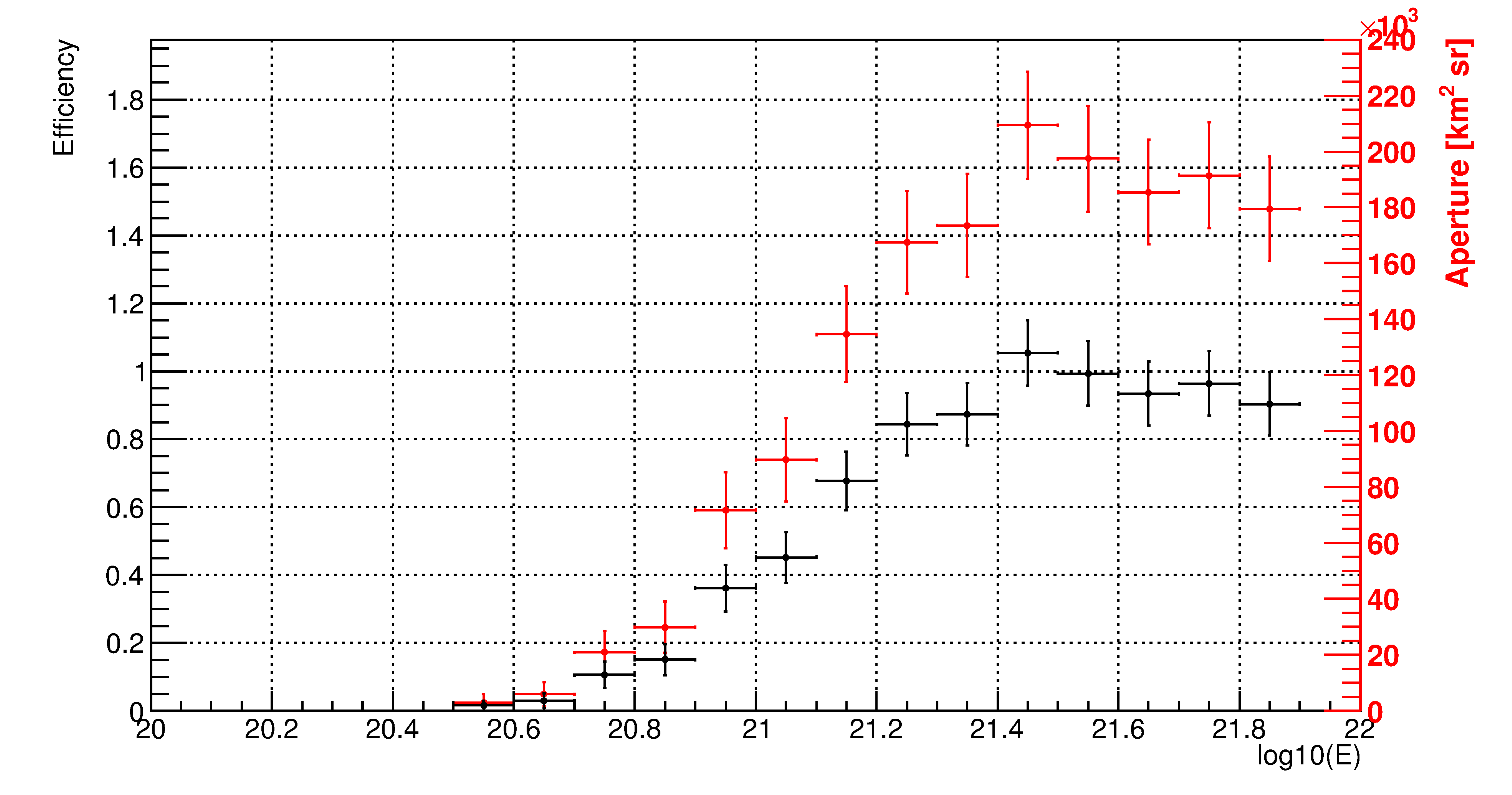}
  \caption{The trigger efficiency of Mini-EUSO as a function of the energy of the air shower, E, is shown in black. The threshold is around $E_{th} = 10^{21}$~\si{\electronvolt}. The effective aperture as a function of energy is also shown in red. }
  \label{fig:eecr_eff}
 \end{figure}

Figure~\ref{fig:eecr} shows the expected track (left) and light curve (right) of a EECR with energy $E$ = \SI{1e21}{\electronvolt}. Figure~\ref{fig:eecr_eff} shows the trigger efficiency curve for Mini-EUSO when adopting the L1 trigger logic described in section \ref{sec:Trigger}.

Despite its energy threshold being too high for cosmic ray detection \\ ($E_{thr}\sim$\SI{1e21}{\electronvolt}), with its annual exposure of $\sim$~\SI{15000}{\square\kilo\metre}~y~\si{\steradian}, Mini-EUSO will provide a significant contribution in estimating an absolute limit on the cosmic ray flux above such energies for a null detection.

As ESAF allows the simulation of phenomena of longer durations such as TLEs, meteors, cities, etc., a few examples of these classes of events were generated in ESAF. These simulations confirmed the capability of the L1 trigger logic to avoid triggers on meteors and cities, whilst in the case of TLEs it was verified that the L1 would trigger if the rising phase of the light curve is so steep that the adaptation of the trigger thresholds at steps of \SI{320}{\micro\second} is still too slow to follow the light increase. Even though the detection of TLEs and lightning is one of the main objectives of the L2 trigger logic, the L1 will allow the recording of the raising phase of the brightest and fastest signals with much higher time resolution.

An estimation of the rate of false triggers due to lower energy galactic cosmic rays hitting the PMTs directly has not yet been taken into account. The effects are currently being investigated using both dedicated simulations of the MAPMT structure and data from the recent EUSO-SPB flight \citep{Wiencke:2016uf}, where such events were detected. From the experience of the TUS instrument \citep{Zotov:2017vh}, these events are indeed expected, and whilst they can occur on similar \si{\micro\second} timescales to those of EECR-like signals, their resulting light profile is sufficiently different such that false triggering can be blocked by a simple trigger veto.

\subsection{L2 trigger tests}
\label{l2_test_esaf}
As described in section \ref{sec:Trigger}, the L2 trigger operates on integrated packets of 128 GTU generated by the L1 trigger. Triggering is performed on the timescale of $\sim$~\SI{40}{\milli\second} with a time resolution of  \SI{320}{\micro\second}, designed to capture the range of transient luminous events (TLEs) in the Earth's atmosphere that will be visible to Mini-EUSO. TLEs are important to study, as they are part of the UV background that will be encountered by future instruments looking to study EECRs from space. However, the high temporal and spatial resolution of Mini-EUSO means that it will also be possible to make unique observations of these atmospheric events, complementing those of other dedicated instruments scheduled to fly in Earth orbit during the same period (e.g. TUS \citep{Panasyuk:2010gk}, ASIM \citep{Neubert:2009ih}).

\begin{figure}[h]
  \centering
  \makebox[\textwidth][c]{
  	\includegraphics[width=0.74\textwidth]{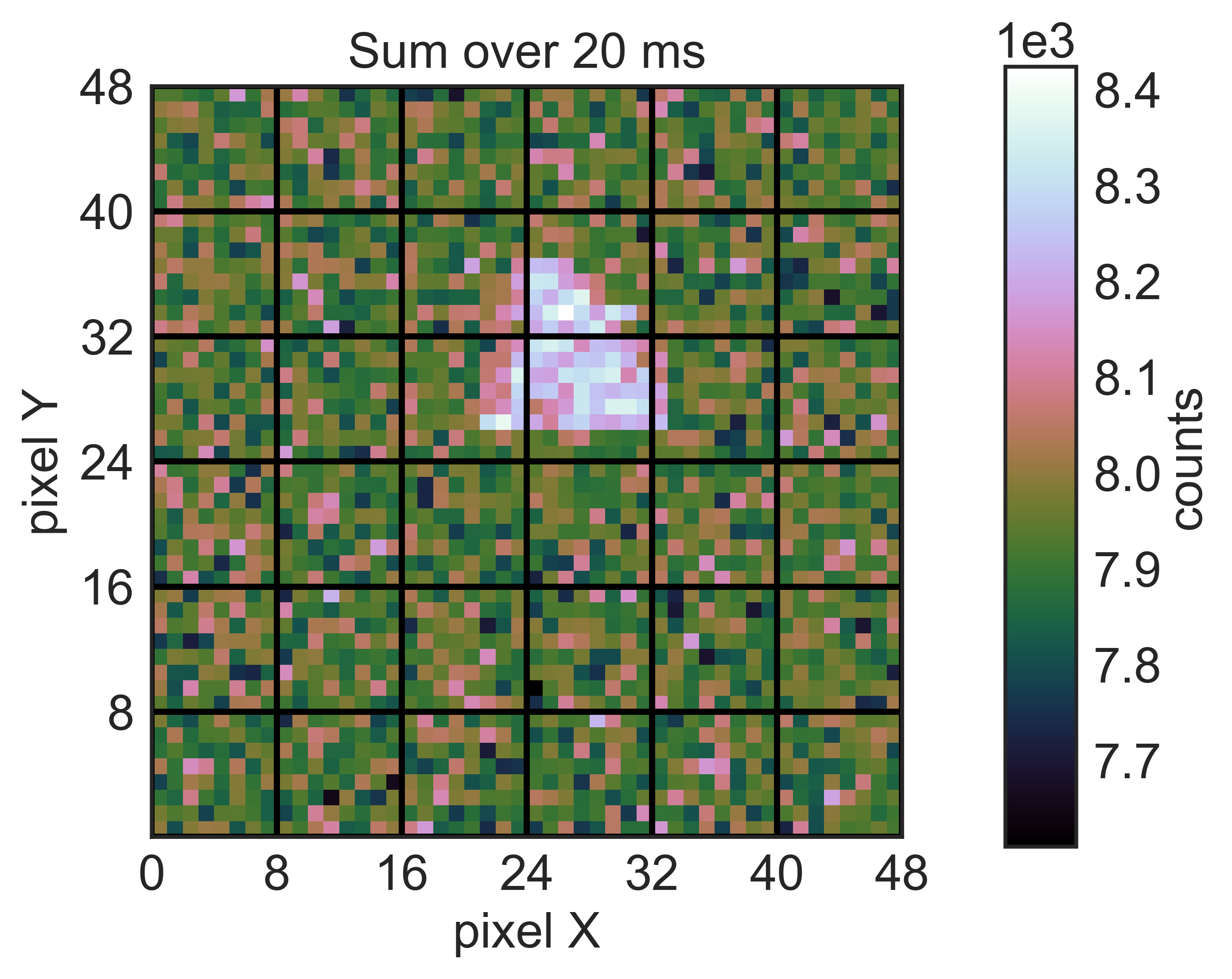} 
    	\includegraphics[width=0.76\textwidth]{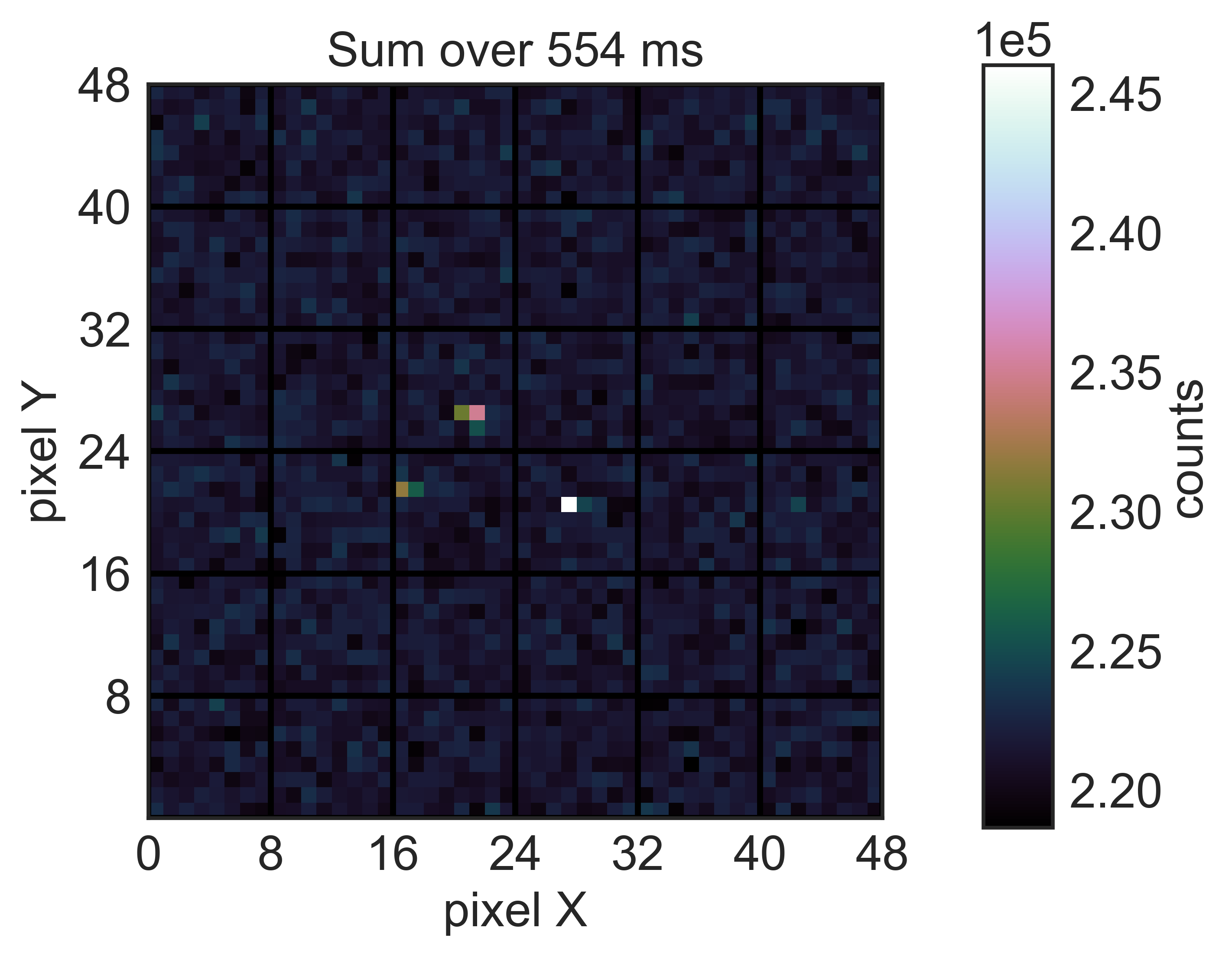}}
  \caption{Left: A typical integrated frame showing a diffuse elf event which brightens the whole PDM.  Right: A similar integrated frame showing 3 localised blue jet events summed over \SI{554}{\milli\second}. The x and y axes represent the pixel grid of the Mini-EUSO focal surface and the colormap shows the number of photons counted by each pixel. All events are simulated with Poissonian background centred on 1 photon/pixel/GTU. Figure taken from \citet{Mini-EUSO}.}
  \label{fig:l2test_event}
 \end{figure}

In order to test the algorithm, the ESAF simulation software was used to generate a range of typical TLE events (namely blue jets, elves and sprites), as would be seen by the Mini-EUSO focal surface. Background was superimposed onto the simulated data packets, assuming a Poisson distribution of background events centred on 1 photon/pixel/GTU \citep{AdamsJr:2013bd}. Examples of the TLE events considered are shown in Figure \ref{fig:l2test_event}. The L2 trigger was then run on this simulated data to test its performance. Two key parameters, the threshold level and the persistence, were varied to investigate their effect on the trigger efficiency. The threshold is simply the level at which the signal is triggered and the persistence is the time frame used to compare the instantaneous signal to background.

\begin{figure}[h]
  \centering
  \includegraphics[width=\textwidth]{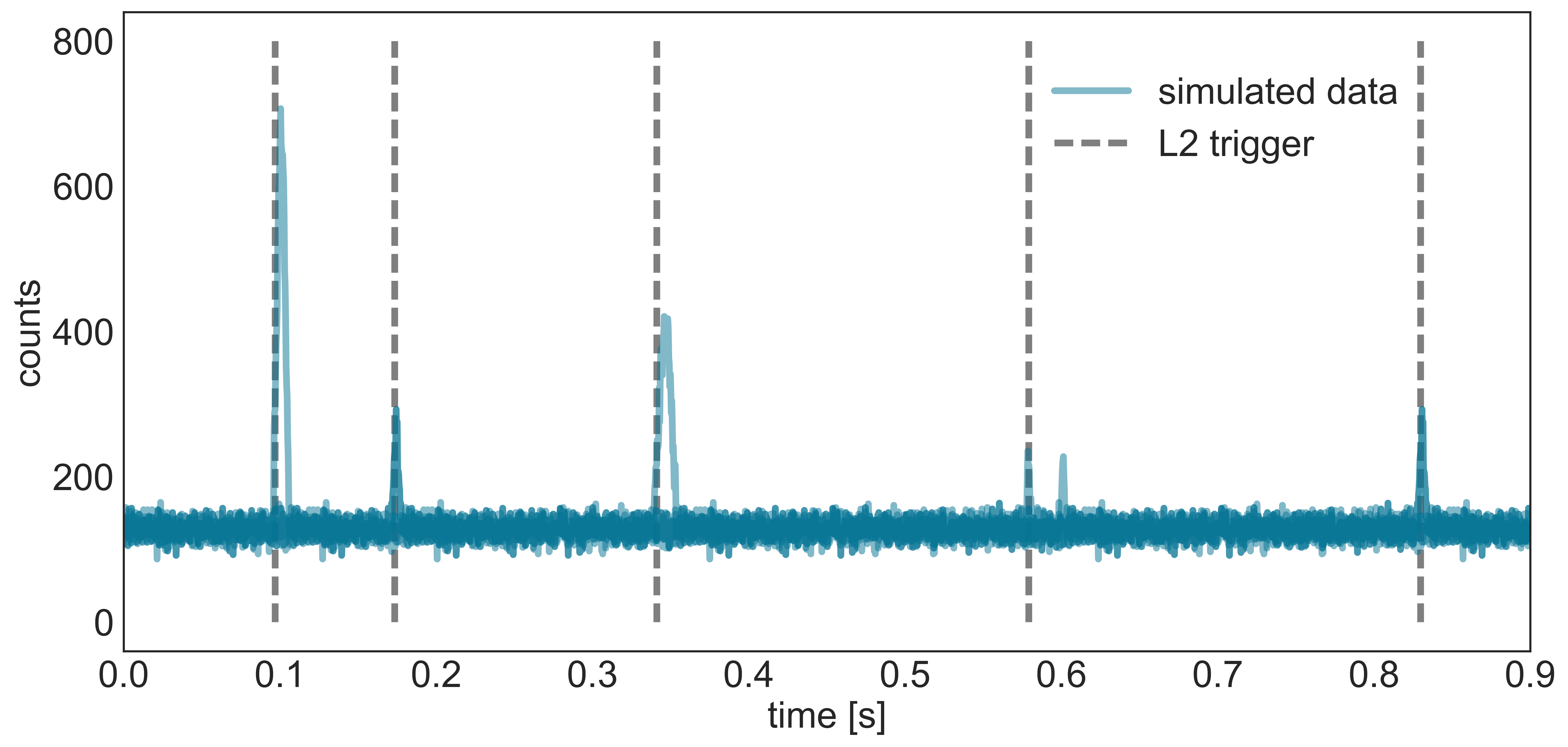}
  \caption{The result of running the L2 trigger algorithm on the simulated TLE. The included events (from left to right) are a blue jet, a sprite, another blue jet, an elf and a final sprite. The events were spread over different areas of the focal surface, and the triggered light curves shown are for a single pixel. The grey dashed line marks an L2 trigger. This result was achieved following the optimisation of the trigger parameters and a trigger efficiency of 100\% is seen.}
  \label{fig:l2test_result}
 \end{figure}

The testing of the trigger algorithm confirmed its ability to distinguish events of interest from typical background levels and also allowed approximate lower limits to be set on the magnitude of the TLEs that Mini-EUSO will be able to detect (for typical sprites and blue jets, an absolute magnitude of $\sim$~3, and for elves an absolute magnitude of $\sim$~1). Figure \ref{fig:l2test_result} shows the trigger response to five different simulated TLEs. The trigger performs well for a threshold of greater than~4~$\times$~the background level and a persistence of 8 \SI{320}{\micro\second} frames. It should be noted that a longer persistence increases the sensitivity of the algorithm to the more diffuse elves, but at the expense of the detection of the more localised blue jets and sprites. The final implementation of the L2 trigger should allow for some compromise here and ideally have parameters which are adjustable in-flight. 

\FloatBarrier
\section{Trigger implementation}
\label{sec: Zoom IN}
\subsection{First tests}
\label{l1_test_hw}
The Mini-EUSO trigger algorithm is coded in VHDL. Following testbench and synthesis simulations, the code was implemented in the PL of the Zynq board and subsequently tested using a pulse generator to induce L1 trigger events. This allowed to test the data acquisition chain from the EC ASIC board to the L1 trigger in the Zynq board of the PDM-DP system. 

In order to do this, a pulse generator was connected via a kapton cable to the MAPMT interface on the EC ASIC board, set to generate an input pulse of \SI{100}{\milli\volt} ($\approx$ 2 photo-electron charge equivalent, for a $5\times10^6$ PMT gain) with a duration of \SI{8}{\nano\second}. This was then connected directly to the PDM-DP system, made up of the cross board, the Zynq board and the power board. The trigger algorithm was programmed to give a L1 event signal to an output pin of the Zynq board upon triggering. This L1 event signal was measured using an oscilloscope. The input pulse amplitude and shape were chosen to simply verify the correct operation of the trigger algorithm in terms of signal over threshold, and not to correspond to a typical signal expected to be detected during the Mini-EUSO mission. As shown in Table \ref{tab:pulsetestres}, the time between pulses was fixed at \SI{100}{\nano\second}.

The number of pulses was varied for a burst rate of \SI{1}{\hertz} and a constant threshold value on the first version of the EC ASIC board (SPACIROC), prior to the upgrade to the SPACIROC 3 ASICs. The main improvements present in the SPACIROC3 are reduced power consumption,  improved double pulse separation and a larger charge dynamic range \cite{BlinBondil:2014ve}, thus the results presented here are also valid for SPACIROC3, which will be used in Mini-EUSO. As the electrical noise was set to $\sim$~1 count/pixel/GTU, the L1 logic presented in Section \ref{sec:Trigger} is expected to start triggering with high efficiency above $\sim$~30 pulses. This is indeed confirmed by the results shown in Table \ref{tab:pulsetestres}. 

Following these initial verification tests, more sophisticated tests of the trigger performance in hardware are currently under way making use of both simulated data and data from the recent flight of EUSO-SPB \citep{Wiencke:2016uf} passed directly into the front-end electronics.

\begin{table}[h]
\centering
\caption{Table showing the results of the pulse generator tests of the L1 trigger. Measurement was taken at a constant DAC level in the SPACIROC1 ASIC, with fixed intervals of \SI{100}{\nano\second} between pulses and a burst rate of \SI{1}{\hertz}. The efficiency of 102\% is due to statistical fluctuations in the background for which the average value was 1 count/pixel/GTU.}
\vspace{3mm}
  \begin{tabular}{ llll }
    \toprule
    \textbf{No. of pulses} & \textbf{No. of trigger/min} & \textbf{Trigger efficiency [\%]} & \textbf{Burst [\si{\micro\second}}] \\ \midrule
    40 & 61 & 102 & 4    \\
    38 & 60 & 100 & 3.8 \\
    36 & 60 & 100 & 3.6 \\
    34 & 42 & 70   & 3.4 \\
    32 & 37 & 62   & 3.2 \\
    30 & 37 & 62   & 3    \\
    20 & 10.3 & 17 & 2 \\
        \bottomrule
  \end{tabular}
  \label{tab:pulsetestres}
\end{table}

\subsection{Ancillary trigger elements}
\begin{figure}[h]
\centering%
{\includegraphics [width=0.75\textwidth]{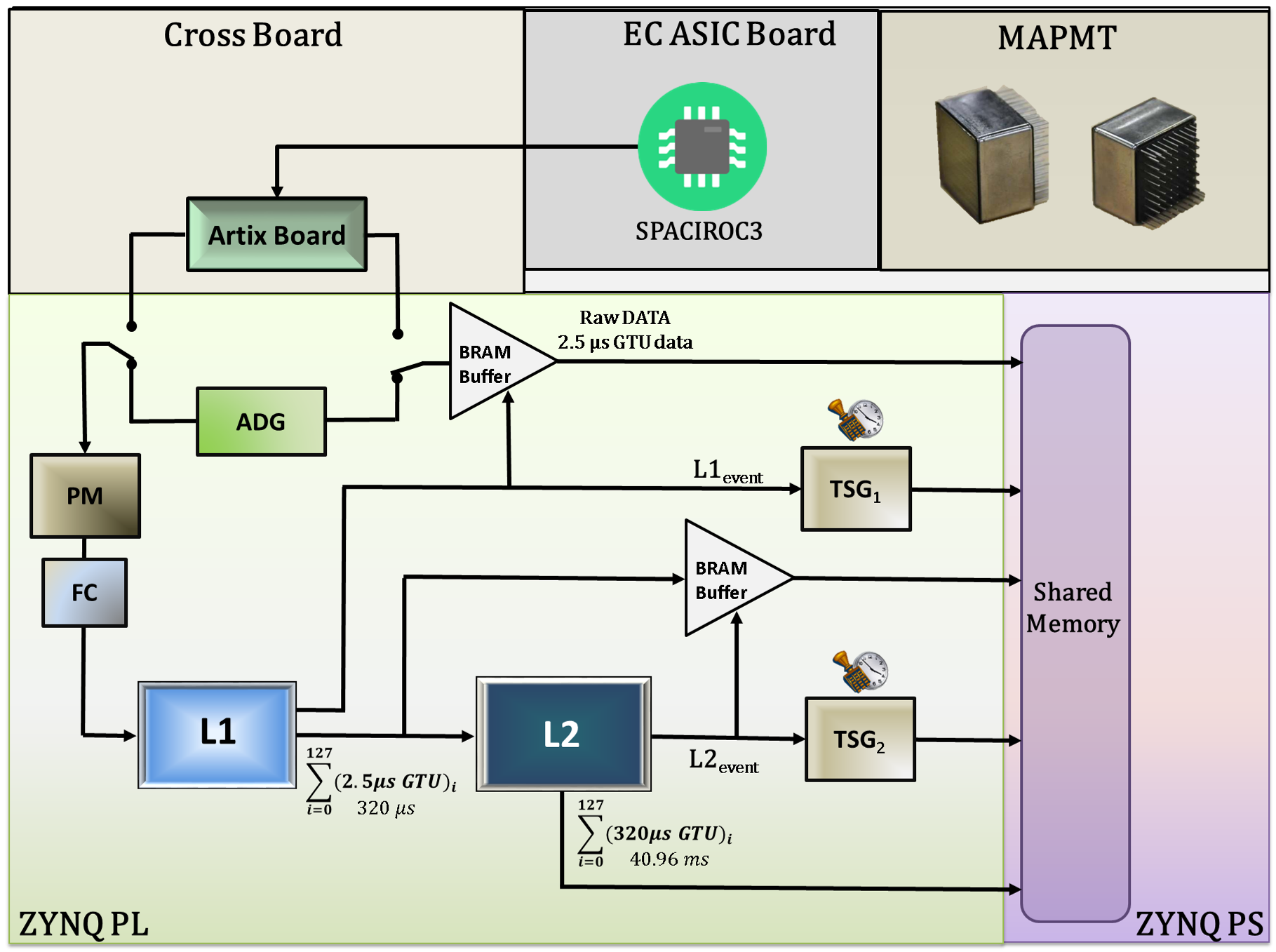}}
\caption{Schematic representation of the trigger logic surroundings. ADG: artificial data generator, PM: pixel masking, FC: format converter, L1: level 1 trigger, L2: level 2 trigger, TSG: time stamp generator.}
\label{fig:surround}
\end{figure}

In addition to the main trigger logic, the artificial data generator, pixel masking module and time stamp generator were also developed and implemented in the PL of the Zynq board, as shown in Figure \ref{fig:surround}. The artificial data generator allows the generation of realistic trigger stimuli within the Zynq board, providing a useful stand-alone testing system for the trigger logic that can easily be used without the main instrument subsystems. It provides both L1 and L2 modes in order to fully test the trigger logic. Pixel masking is implemented in order to mask pixels showing unexpected behaviour. This is important in order to control fake triggers an maximise the scientific return of the instrument. The pixel masking interfaces to the PS of the Zynq and the Mini-EUSO CPU, to allow pixels to be masked in-flight via the uploading of a configuration file.

The time stamp generator is needed in order to tag triggered events precisely. Upon boot, the Zynq board is synchronised with the Mini-EUSO CPU and counts the incoming data at the GTU level. When a trigger occurs, the corresponding GTU number is read out with the event and this information is passed back to the CPU. 
 
\FloatBarrier
\section{Conclusion}
 \label{sec:conclusion}
The Mini-EUSO trigger algorithm has been integrated in the Zynq Board FPGA. Prior to this, the trigger algorithm was tested successfully using simulated data and data generated as part of the EUSO@TurLab project. Once integrated in the hardware, the trigger was then tested using a pulse generator and the complete data acquisition chain. The artificial data generator implemented in the Zynq board will allow for stand alone testing of the trigger logic. Following the trigger implementation in the PDM-DP system will now be integrated with the Mini-EUSO instrument, allowing for end-to-end testing of the data acquisition system. 

\section*{Acknowledgments}
This work was partially supported by the Italian Ministry of Foreign Affairs and International Cooperation, Italian Space Agency (ASI) contract 2016-1-U.0, the Russian Foundation for Basic Research, grants \#15-35-21038 and \# 16-29-13065, and the Olle Engkvist Byggm{\"a}stare Foundation. The technical support of G.~Cotto, R.~Forza, and M.~Manfrin during validation tests at TurLab is deeply acknowledged.

\section*{References}

\bibliography{mybibfile}

\begin{thebibliography}{19}
\providecommand{\natexlab}[1]{#1}
\providecommand{\url}[1]{\texttt{#1}}
\expandafter\ifx\csname urlstyle\endcsname\relax
  \providecommand{\doi}[1]{doi: #1}\else
  \providecommand{\doi}{doi: \begingroup \urlstyle{rm}\Url}\fi

\bibitem[Abraham et~al.(2004)Abraham, Aglietta, Aguirre,
  et~al.]{Abraham:2004dt}
J.~Abraham, M.~Aglietta, I.~C. Aguirre, et~al.
\newblock {Properties and performance of the prototype instrument for the
  Pierre Auger Observatory}.
\newblock \emph{Nucl. Instrum. Meth.}, A523\penalty0 (1-2):\penalty0 50--95,
  2004.

\bibitem[Berat et~al.(2010)Berat, Bottai, De~Marco, et~al.]{Berat10}
C.~Berat, S.~Bottai, D.~De~Marco, et~al.
\newblock {Full simulation of space-based extensive air showers detectors with
  ESAF}.
\newblock \emph{Astroparticle Physics}, 33\penalty0 (4):\penalty0 221--247, May
  2010.

\bibitem[Bertaina et~al.(2014)Bertaina, Biktemerova, Bittermann,
  et~al.]{Bertaina:2014fk}
M.~Bertaina, S.~Biktemerova, K.~Bittermann, et~al.
\newblock {Performance and air-shower reconstruction techniques for the
  JEM-EUSO mission}.
\newblock \emph{Advances in Space Research}, 53\penalty0 (10):\penalty0
  1515--1535, May 2014.

\bibitem[Bertaina et~al.(2015)Bertaina, Bowaire, Cambursano,
  et~al.]{Bertaina:2015hha}
M.~Bertaina, A.~Bowaire, S.~Cambursano, et~al.
\newblock {EUSO@TurLab: An experimental replica of ISS orbits}.
\newblock \emph{EPJ Web of Conferences}, 89:\penalty0 03003, Mar. 2015.

\bibitem[Blin-Bondil et~al.(2014)Blin-Bondil, Barrillon, Dagoret-Campagne,
  et~al.]{BlinBondil:2014ve}
S.~Blin-Bondil, P.~Barrillon, S.~Dagoret-Campagne, et~al.
\newblock {SPACIROC3: A Front-End Readout ASIC for JEM-EUSO cosmic ray
  observatory}.
\newblock In \emph{Proceedings of the 3rd International Conference on
  Technology and Instrumentation in Particle Physics (TIPP), Amsterdam}, 2014.
\newblock PoS(TIPP2014)172.

\bibitem[Capel et~al.(2017)Capel, Belov, Casolino, et~al.]{Mini-EUSO}
F.~Capel, A.~Belov, M.~Casolino, et~al.
\newblock {Mini-EUSO: A high resolution detector for the study of terrestrial
  and cosmic UV emission from the International Space Station}.
\newblock \emph{Advances in Space Research}, 2017.
\newblock 10.1016/j.asr.2017.08.030.

\bibitem[Ebisuzaki et~al.(2014)Ebisuzaki, Medina-Tanco, and
  Santangelo]{Ebisuzaki:2014wka}
T.~Ebisuzaki, G.~Medina-Tanco, and A.~Santangelo.
\newblock {The JEM-EUSO mission}.
\newblock \emph{Adv. Space Res.}, 53\penalty0 (10):\penalty0 1499--1505, 2014.

\bibitem[Franz et~al.(1990)Franz, Nemzek, and Winckler]{Franz:1990cu}
R.~C. Franz, R.~J. Nemzek, and J.~R. Winckler.
\newblock {Television Image of a Large Upward Electrical Discharge Above a
  Thunderstorm System}.
\newblock \emph{Science}, 249\penalty0 (4964):\penalty0 48--51, July 1990.

\bibitem[Hachisu et~al.(2011)Hachisu, Tone, Uehara, et~al.]{Hachisu:2011vx}
Y.~Hachisu, N.~Tone, Y.~Uehara, et~al.
\newblock {JEM-EUSO lens manufacturing}.
\newblock In \emph{Proceedings of the 32nd International Cosmic Ray Conference,
  Beijing}, volume~3, pages 184--187, 2011.

\bibitem[Kawai et~al.(2008)Kawai, Yoshida, Yoshii, Tanaka,
  et~al.]{Kawai:2008gz}
H.~Kawai, S.~Yoshida, H.~Yoshii, K.~Tanaka, et~al.
\newblock {Telescope Array Experiment}.
\newblock \emph{Nuclear Physics B - Proceedings Supplements}, 175-176:\penalty0
  221--226, Jan. 2008.

\bibitem[Neubert and {{the ASIM Instrument Team}}(2009)]{Neubert:2009ih}
T.~Neubert and {{the ASIM Instrument Team}}.
\newblock {ASIM{\textemdash}an Instrument Suite for the International Space
  Station}.
\newblock \emph{{Coupling of thunderstorms and lightning discharges to
  near-Earth space, Corte (France)}}, 1118\penalty0 (1):\penalty0 8--12, May
  2009.

\bibitem[Olinto et~al.(2015)Olinto, Parizot, Bertaina, and
  Medina-Tanco]{Olinto15}
A.~V. Olinto, E.~Parizot, M.~Bertaina, and G.~Medina-Tanco.
\newblock {JEM-EUSO Science}.
\newblock In \emph{{Proceedings of the 34th International Cosmic Ray
  Conference}}, 2015.
\newblock PoS(ICRC2015)623.

\bibitem[Panasyuk et~al.(2015)Panasyuk, Picozza, and Casolino]{Panasyuk15}
M.~Panasyuk, P.~Picozza, and M.~Casolino.
\newblock {Ultra high energy cosmic ray detector (KLYPVE) on board the Russian
  Segment of the ISS}.
\newblock In \emph{{Proceedings of the 34th International Cosmic Ray
  Conference}}, 2015.
\newblock PoS(ICRC2015)669.

\bibitem[Panasyuk et~al.(2010)Panasyuk, Bogomolov, Garipov,
  et~al.]{Panasyuk:2010gk}
M.~I. Panasyuk, V.~V. Bogomolov, G.~K. Garipov, et~al.
\newblock {Transient luminous event phenomena and energetic particles impacting
  the upper atmosphere: Russian space experiment programs}.
\newblock \emph{Journal of Geophysical Research: Space Physics}, 115\penalty0
  (A6), June 2010.

\bibitem[Pasko et~al.(2011)Pasko, Yair, and Kuo]{Pasko:2011ev}
V.~P. Pasko, Y.~Yair, and C.-L. Kuo.
\newblock {Lightning Related Transient Luminous Events at High Altitude in the
  Earth{\textquoteright}s Atmosphere: Phenomenology, Mechanisms and Effects}.
\newblock \emph{Space Science Reviews}, 168\penalty0 (1-4):\penalty0 475--516,
  Sept. 2011.

\bibitem[{The JEM-EUSO Collaboration}(2013)]{AdamsJr:2013bd}
{The JEM-EUSO Collaboration}.
\newblock {An evaluation of the exposure in nadir observation of the JEM-EUSO
  mission}.
\newblock \emph{Astroparticle Physics}, 44:\penalty0 76--90, 2013.

\bibitem[{The JEM-EUSO Collaboration}(2015)]{Collaboration15}
{The JEM-EUSO Collaboration}.
\newblock {The JEM-EUSO instrument}.
\newblock \emph{Experimental Astronomy}, 40\penalty0 (1):\penalty0 19--44,
  2015.

\bibitem[Wiencke and {{the JEM-EUSO collaboration}}(2015)]{Wiencke:2016uf}
L.~Wiencke and {{the JEM-EUSO collaboration}}.
\newblock {EUSO-Balloon mission to record extensive air showers from near
  space}.
\newblock In \emph{Proceedings of the 34th International Cosmic Ray
  Conference}, 2015.
\newblock PoS(ICRC2015)631.

\bibitem[Zotov and {{the Lomonosov-UHECR TLE
  collaboration}}(2017)]{Zotov:2017vh}
M.~Zotov and {{the Lomonosov-UHECR TLE collaboration}}.
\newblock {Early Results from TUS, the First Orbital Detector of Extreme Energy
  Cosmic Rays}.
\newblock \emph{arXiv.org/astro-ph.IM}, Mar. 2017.
\newblock 1703.09484v3.

\end{thebibliography}

\end{document}